\documentclass[aps,prl,nofootinbib,amsmath,twocolumn,preprintnumbers]{revtex4-1}
\usepackage{amssymb,esvect,amsmath,graphicx,latexsym,amsthm,slashed,eso-pic,hyperref}
\usepackage{float}
\usepackage[hang,flushmargin]{footmisc}
\usepackage{tikzfeynman}
\usepackage{placeins}
\usepackage{bbold}
\usepackage{multirow}
\usepackage{slashed}
\usepackage{cancel}

\newcommand{\beq}{\begin{equation}} \newcommand{\eeq}{\end{equation}}
\newcommand{\bea}{\begin{eqnarray}} \newcommand{\eea}{\end{eqnarray}}

\DeclareMathOperator{\br}{Br}

\begin{document}
\preprint{YITP-SB-19-43}

\title{Light Scalars and the KOTO Anomaly}

\author{Daniel Egana-Ugrinovic$^{1}$}
\author{Samuel Homiller$^{2}$}
\author{Patrick Meade$^{2}$}
\affiliation{
$^{1}$Perimeter Institute for Theoretical Physics, Waterloo, ON N2L 2Y5\\
$^{2}$C. N. Yang Institute for Theoretical Physics, Stony Brook University, 
 Stony Brook, NY 11794}

\begin{abstract}
The KOTO experiment recently presented a significant excess of events in their search for the rare SM process $K_L \to \pi^0\nu\bar{\nu}$, well above both Standard Model signal and background predictions.  We show that this excess may be due to weakly-coupled scalars that are produced from Kaon decays and escape KOTO undetected.
We study two concrete realizations, the minimal Higgs portal and a hadrophilic scalar model,
and demonstrate that they can explain the observed events while satisfying bounds from other flavor and beam-dump experiments.  Hadronic beam-dump experiments provide particularly interesting constraints on these types of models, and we discuss in detail the normally underestimated uncertainties associated with them.
The simplicity of the models which can explain the excess, and their possible relations with interesting UV constructions, 
provides strong theoretical motivation for a new physics interpretation of the KOTO data.

\end{abstract}

\maketitle

%%%%%%%%%%%%%%%%%%%%%%%%%%%%%%%%%%%%%%%%%%%%%%%%%%%%%%%%
%%%%%%%%%%%%%%%%%%%%%%%%%%%%%%%%%%%%%%%%%%%%%%%%%%%%%%%%
\section{Introduction} 

Recently, the KOTO experiment presented an excess of events compared to the SM expectation
in the signal region for the rare process $K_L \to \pi^0\nu\bar{\nu}$~\cite{KOTO:2019}.
Four events were observed,
compared to a Standard Model (SM) plus background expectation of only $0.1\pm 0.02$ events.
\footnote{The $\mathcal{O}(10\%)$ systematics in the event sensitivity are  neglected here \cite{Ahn:2018mvc}.}
One event is believed to originate from SM activity upstream from the detector, 
but the remaining three are currently unexplained.
Given the experiment's branching fraction sensitivity of $6.9\times 10^{-10}$ for single events, 
the three events are consistent with  
%boe%
\begin{equation}
\br(K_L \to \pi^0 \nu \bar{\nu})_{\mathrm{KOTO}} = 2.1^{+2.0 (+4.1)}_{-1.1 (-1.7)} \times 10^{-9} \quad ,
\label{eq:KOTOresult}
\end{equation}
%eoe%
where the uncertainties are due to statistics.
This result is two-orders of magnitude larger than the SM prediction, 
$\br(K_L \to \pi^0 \nu\bar{\nu})_{\mathrm{SM}} = (3.4 \pm 0.6) \times 10^{-11}$%, 
\cite{Cirigliano:2011ny}.

In this letter, 
we study a simple new physics interpretation of the KOTO results.
We focus on the possibility that the excess is due to new decays, $K_L \rightarrow \pi^0 \varphi$, where $\varphi$ is a light new scalar ~\cite{Hou:2016den,Kitahara:2019lws}, 
which is sufficiently long-lived and weakly interacting so that it appears as missing energy at KOTO.

We study two concrete realizations. 
First, we consider the real-scalar singlet extension of the SM \cite{Veltman:1989vw,McDonald:1993ex,OConnell:2006rsp,Egana-Ugrinovic:2015vgy},
also referred to as the minimal ``Higgs portal".
This model exemplifies simplicity and minimality: 
it is the most trivial extension of the SM, 
and does not require any new flavor structure beyond the SM.
In addition, the minimal Higgs portal shares many similarities with models with light dilatons or radions \cite{Gildener:1976ih,Goldberger:1999uk,Goldberger:2008zz,Damour:2010rp,Bellazzini:2012vz,Abe:2012eu,Chacko:2012sy,Coradeschi:2013gda},
which arise in well-motivated UV constructions.
As a second possibility, we study hadrophilic scalar models with flavor-aligned, generation-specific couplings.
These models arise naturally in UV constructions where flavor mixing is due to quark-wave function renormalization, 
which are referred to a spontaneous flavor-violating (SFV) \cite{Egana-Ugrinovic:2018znw,Egana-Ugrinovic:2019dqu}.
Hadrophilic models with couplings to up-type quarks lead to $K_L\to \pi^0 \varphi$ at one loop as the minimal Higgs portal, 
and in addition, they allow us to explore the possibility that the KOTO excess is due to novel flavor textures, which may lead to different kinematics.

The challenge for new physics interpretations of the KOTO results is to explain the large number of observed events,
while being consistent with strong bounds from beam-dump experiments, charged Kaon factories and other flavor experiments. 
While models with new light scalars are strongly constrained \cite{Beacham:2019nyx}, 
we show that in both our concrete realizations there are open regions of parameter space consistent with the KOTO results.
Our selection of models allows us to illustrate gaps in bounds from charged Kaon factories that have been pointed out in the literature \cite{Hou:2016den,Kitahara:2019lws}, 
and to find a few small open regions where searches need to be improved. 
Through a careful analysis of existing bounds, 
we find that a minimal Higgs portal with scalar mass in the range $110 \, \textrm{MeV} \leq m_{\varphi} \leq 180 \, \textrm{MeV}$ 
may be the origin of the KOTO excess.
We also find small regions of parameter space for masses below $\leq 60 \, \textrm{MeV}$, 
which are consistent with the excess.
For a hadrophilic, flavor-aligned scalar models coupling preferentially to up-type quarks, 
which arise in SFV theories, 
we can categorize models assuming a dominant coupling to $u,c,$ or $t$ quarks.
We find that top-philic scalars lead to similar conclusions as in the minimal Higgs portal.
For charm-philic scalars,
we find agreement with the KOTO excess for the range $100 \, \textrm{MeV}  \leq  m_{\varphi} \leq 180 \, \textrm{MeV}$,
but also some tension with bounds from beam dump experiments. 
Finally, for a singlet coupling mostly to the up quark, 
we find no consistent interpretation of the KOTO results. 

Our analysis demonstrates that \textit{extremely simple} and \textit{motivated} models of new physics,
especially the minimal Higgs portal and phenomenologically similar constructions,
may be the origin of the KOTO excess. 
From a theoretical perspective, 
this provides strong support for a new physics explanation of the announced results.

We organize this letter as follows. 
In the first section, we present the minimal Higgs portal and hadrophilic scalar models.
In the second and third sections, 
we study the KOTO excess in the context of each one of these models, respectively.
We conclude with UV motivations for our models and comment on experimental signatures that could test our scenario. 
In a first appendix we include a discussion of hadronic beam dump experiments and their uncertainties, 
which are relevant to general BSM models that can be probed by these experiments. 
In a second appendix we obtain the rate of $\eta$ decays into a pion and a scalar, required to compute scalar production rates at beam dumps,
and compare the results with previous calculations available in the literature.

%%%%%%%%%%%%%%%%%%%%%%%%%%%%%%%%%%%%%%%%%%%%%%%%%%%%%%%%
%%%%%%%%%%%%%%%%%%%%%%%%%%%%%%%%%%%%%%%%%%%%%%%%%%%%%%%%
\section{Scalars with Flavored Couplings}
\subsection{Minimal Higgs Portal}
We extend the SM with a light real scalar singlet $S$. 
At the renormalizable level, the Lagrangian for our singlet and the Higgs is
\begin{eqnarray}
\nonumber
\mathcal{L} 
& \supset &
D_{\mu}H^{\dagger}D^{\mu}H  +
 \frac{1}{2} \partial_\mu S \partial^\mu S  -  V(S,H) 
  \\
 &-& \bigg[\lambda^{u}_{ij}Q_i H \bar{u}_j 
- \lambda^{d\dagger}_{ij} Q_i {H}^c \bar{d}_j 
- \lambda^{\ell \dagger}_{ij} L_i {H}^c \bar{\ell}_j  
\bigg]
\quad ,
\label{eq:renL}
\end{eqnarray}
where the potential can be found in \cite{Egana-Ugrinovic:2015vgy}.
In this theory, the singlet and the CP-even neutral scalar in $H$ mix in the mass matrix.
The corresponding two mass eigenstates are the $125 \, \textrm{GeV}$ Higgs boson $h$, 
and a new scalar boson $\varphi$ with mass $m_\varphi$.
 The couplings of the new scalar to SM fields are equal to the Higgs couplings, 
up to a universal mixing angle $\theta$.
In particular, the couplings to fermions in the mass eigenbasis are flavor-diagonal and given by
\begin{equation}
\lambda_{\varphi}^f = - \sin\theta \frac{m_f}{v} \quad ,
\label{eq:higgsportalcouplings}
\end{equation}
where $m_f$ are the SM fermion masses.
\subsection{Hadrophilic Scalar Coupling to Up-type Quarks}
 
The minimal Higgs portal theory constrains the scalar-fermion couplings to follow the SM hierarchies, 
limiting the scalar's flavor phenomenology.
To discuss the flavor structure of our scalar model in more generality,
we now allow for flavor-specific couplings with the SM quarks.  
These couplings can be obtained by going beyond the renormalizable level interactions of Eq.~\eqref{eq:renL}, 
and adding dimension-five scalar-fermion operators. 
Here we limit ourselves to include non-renormalizable interactions between our scalar and up-type quarks only \cite{Batell:2017kty,Batell:2018fqo},
%boe%
\begin{equation}
\label{eq:dim5_yukawas}
\mathcal{L} \supset \frac{S}{M} c^u_{ij} Q_i H \bar{u}_j  \quad ,
\end{equation}
%eoe%
where $M$ points to the scale of the UV completion leading to the dimension-five operator, 
and $c^u_{ij}$ is a new Yukawa matrix, 
which leads to novel flavored interactions.
The operator~\eqref{eq:dim5_yukawas} can be easily obtained in UV completions with an extra Higgs doublet \cite{Egana-Ugrinovic:2019dqu} or vector-like quarks \cite{Batell:2017kty}.
To avoid tree-level FCNC's mediated by the new scalar, 
we impose that $c^u_{ij}$ is simultaneously diagonalizable with the up-type quark SM Yukawa, 
\textit{i.e.}, 
that it is flavor-aligned.
While in the low energy effective theory Eq.~\eqref{eq:dim5_yukawas} there is no evident symmetry to guarantee flavor-alignment, 
in \cite{Egana-Ugrinovic:2018znw,Egana-Ugrinovic:2019dqu} it was shown that this can be imposed by a UV flavor construction called down-type Spontaneous Flavor Violation.

In the limit of vanishing scalar mixing $\theta \rightarrow 0$,
the new scalar is hadrophilic (and leptophobic),
and couples to up-type quarks only due to the non-renormalizable interaction Eq.~\eqref{eq:dim5_yukawas}. 
In the quark mass eigenbasis, these couplings are flavor-diagonal and related to the couplings of the dimension-five operator Eqns.~\eqref{eq:dim5_yukawas} via
%boe%
\begin{equation}
\lambda_{\varphi}^q
= 
v/(\sqrt{2} M) \kappa_{q} 
 \quad ,  \quad q=u,c,t \quad ,
\label{eq:hadrophilicYukawas}
\end{equation}
%eoe%
where $\kappa_{u,c,t}$ are three independent Yukawa couplings controlling the interactions of the singlet to up-type quarks, 
which \textit{do not necessarily follow the SM hierarchies}.  
In particular we can study theories which have a coupling to only one individual up-type quark at a time, 
letting us explore the contributions to the KOTO excess systematically.

%%%%%%%%%%%%%%%%%%%%%%%%%%%%%%%%%%%%%%%%%%%%%%%%%%%%%%%%
%%%%%%%%%%%%%%%%%%%%%%%%%%%%%%%%%%%%%%%%%%%%%%%%%%%%%%%%
\section{Minimal Higgs Portal explanation of the KOTO excess}

%%%%%%%%%%%%%%%%%%%%%%%%%%%%
\begin{figure}
\centering
\includegraphics[scale=0.33]{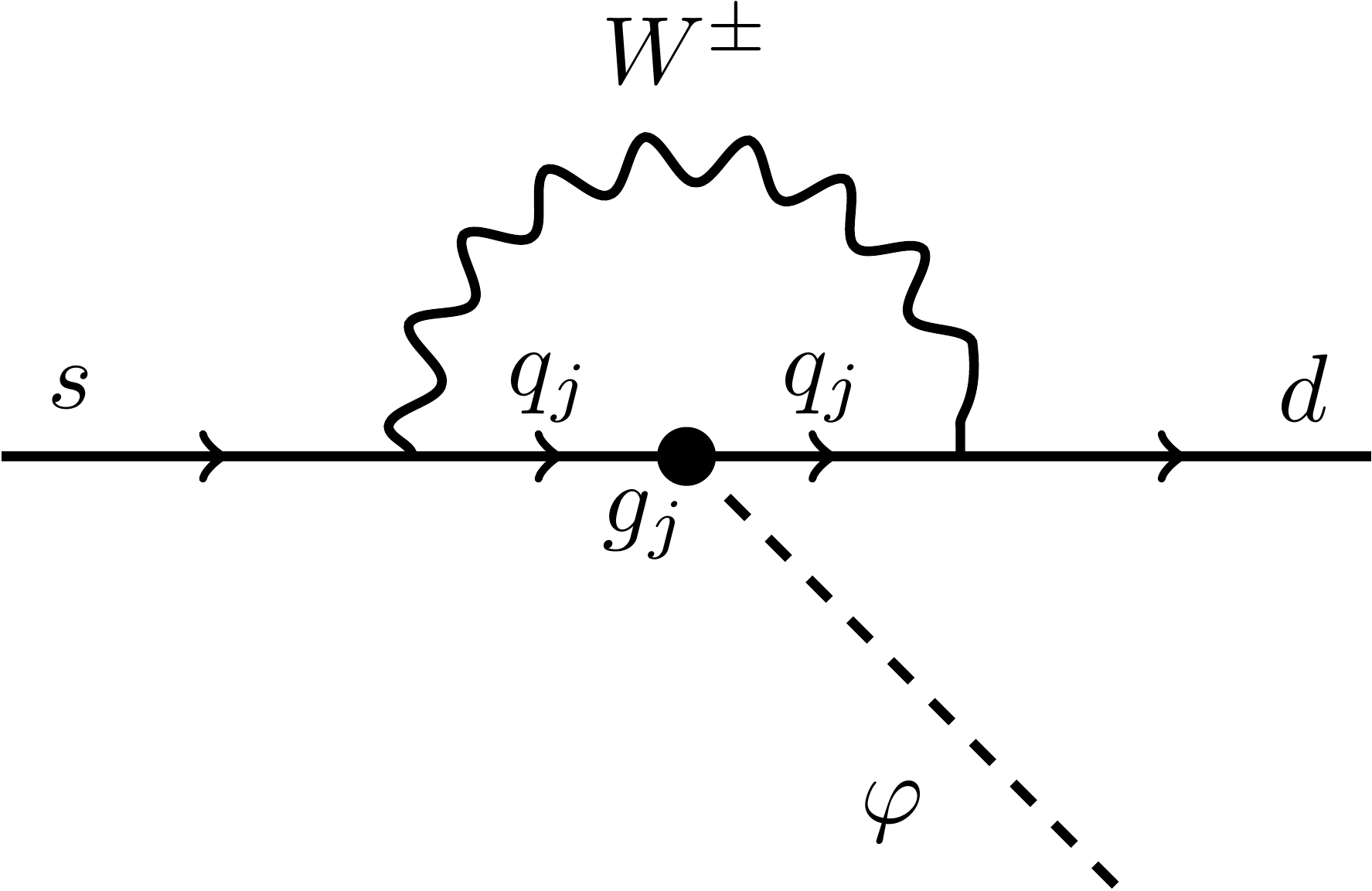}
\caption{Penguin diagram leading to $K \to \pi \varphi$, 
where $\varphi$ is our new scalar particle.}
\label{fig:penguin}
\end{figure}
 %%%%%%%%%%%%%%%%%%%%%%%%%%%%

Light scalars with couplings to quarks are produced at KOTO via penguin diagrams with internal up, charm or top quarks, 
as in Fig. \ref{fig:penguin}.
The corresponding decay width is given by~\cite{Leutwyler:1989xj, Gunion:1989we, Bezrukov:2009yw}
\begin{eqnarray}
\label{eq:neutralkaon}
\nonumber
\Gamma_{K_L \to \pi^0 \varphi}  
&=& 
\frac{(\mathrm{Re}\, g_{\varphi K \pi} )^2}{16\pi m_K^3} \lambda^{1/2}(m_K^2, m_{\pi}^2, m_\varphi^2)  \quad , 
\\
g_{\varphi K \pi} &=&  \frac{3 m_K^2} {32 \pi^2 v^2} 
\sum_{f = u, c, t}
\lambda_{\varphi}^f
\, m_f\,V_{fd}^* V_{fs}^{\phantom{*}} \quad .
\end{eqnarray}
where the scalar couplings $\lambda_{\varphi}^f$ are given in Eq.~\eqref{eq:higgsportalcouplings} and $\lambda$ is the triangle function.
If the new scalar escapes the KOTO detector before decaying into SM fields, 
the event falls into the signal region for $K_L \to \pi^0 \nu \bar{\nu}$, 
and may be the explanation for the observed excess. 
The contribution of the scalars to the effective branching fraction measured at KOTO is
\begin{equation}
\br^{\textrm{eff}}(K_L \to \pi^0 \nu \bar{\nu})  = 
\epsilon
\,
\br(K_L \to \pi^0 \varphi) 
\,
e^{\left(-\frac{m_\varphi}{c\tau_{\varphi}} \frac{ L}{p_{\varphi}}\right)}
\quad ,
\label{eq:effectiveBR}
\end{equation}
%eoe%
where $\br(K_L \to \pi \varphi) $ is obtained from Eq.~\eqref{eq:neutralkaon}, 
the SM Kaon width is $\Gamma_{K_L}^{\mathrm{SM}} = 1.29\times 10^{-17}\,\mathrm{GeV}$ \cite{Patrignani:2016xqp},
the exponential suppression accounts for the scalars that decay before escaping the KOTO detector, 
and $\epsilon$ is a correction factor that accounts for the kinematical difference between the 3-body SM decay process, 
and the 2-body decay into our scalar $\varphi$. 
This factor is taken from \cite{Ahn:2018mvc}, and varies from $\epsilon = 0.75$ for a massless scalar to $\epsilon=1$ for $m_\varphi= 200 \, \textrm{MeV}$.
In the exponential factor, the KOTO detector size is $L \, = \, 3\,\mathrm{m}$ and $p_{\varphi}$ is the scalar's momentum. 
The typical momentum was obtained from a KOTO simulation in \cite{Kitahara:2019lws},
and corresponds to an energy for the scalar particle of  $E_{\varphi} = 1.5\,\mathrm{GeV}$.
The scalar's lifetime $c\tau_{\varphi}$ is completely specified by the mixing angle $\theta$ and its mass  $m_{\varphi}$. 
For our Higgs portal discussion we limit ourselves to masses in the range $2 m_e < m_{\varphi} \leq 200 \, \textrm{MeV}$.
The upper end is motivated by the large transverse momentum of the pions in the observed events at KOTO, 
which translates into an upper bound on the scalar mass \cite{Kitahara:2019lws}.
The lower end of this range is chosen for simplicity:
for this range of masses the scalar's lifetime is controlled mostly by the decay to electrons \cite{Clarke:2013aya}.

%%%%%%%%%%%%%%%%%%%%%%%%%%%%
\begin{figure}[h!]
\includegraphics[width=8cm]{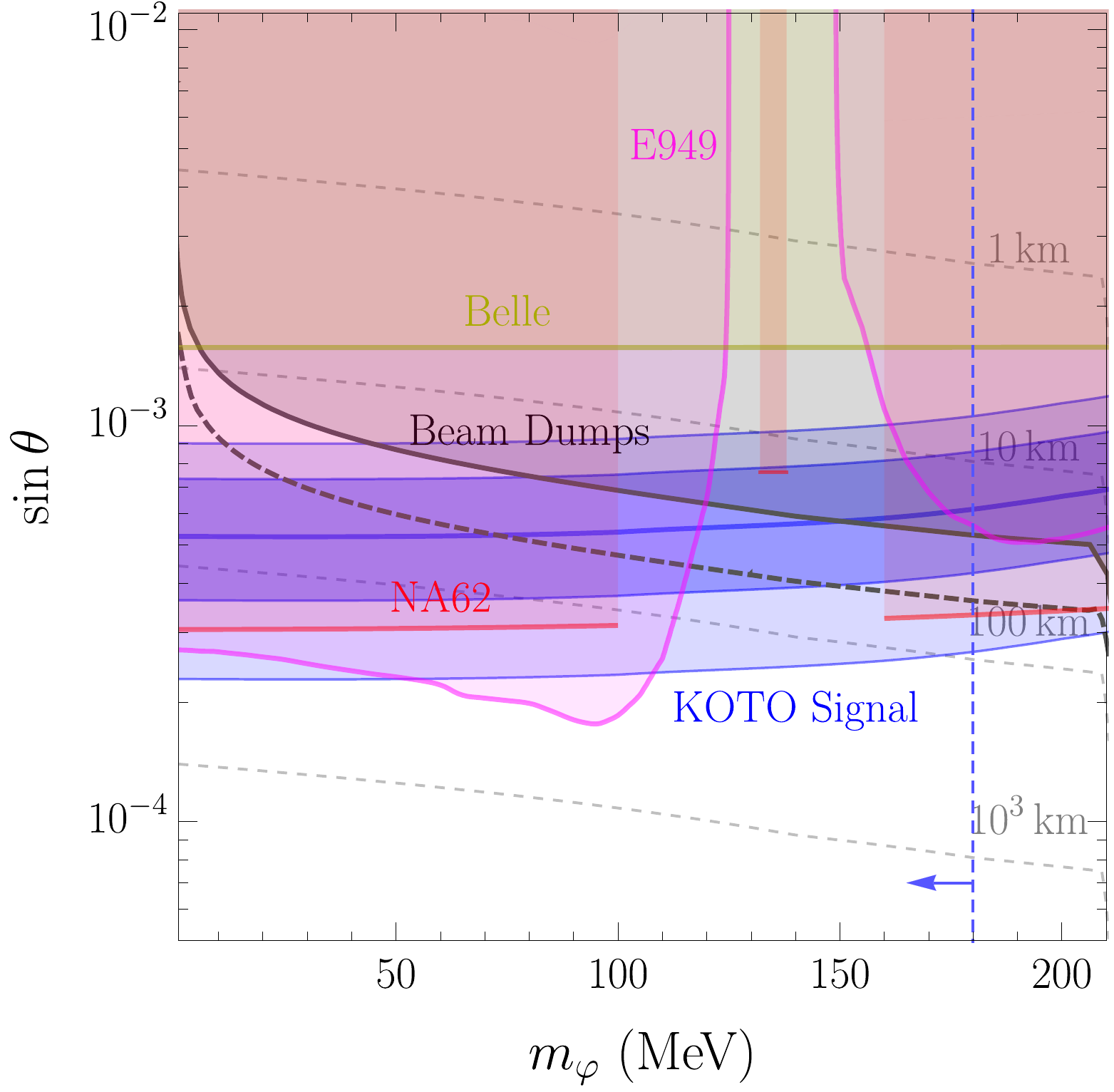}
\caption{
Minimal Higgs portal interpretation of the KOTO excess, 
and leading bounds on the model, 
plotted as a function of the singlet-like scalar mass and mixing angle with the Higgs.
\textit{Blue:} regions of parameter space consistent with the number of $K_L \to \pi^0 \nu \bar{\nu}$ events observed at KOTO.
The solid line corresponds to the measured central value, while the shaded regions include the $1$ and $2\sigma$ compatible values. 
The region to the left of the vertical-dashed blue line corresponds to singlet masses that are consistent with the kinematics of the observed KOTO events.
\textit{Red:} 
limits from NA62 on $\br(K^+ \to \pi^+ \nu \bar{\nu})$, 
and $\br(K^+ \to \pi^+\,\pi^0)$ with $\pi^0$ decaying invisibly.
\textit{Pink:} 
limits from E949 on $\br(K^+ \to \pi^+  X)$ with $X$ a long-lived particle.
\textit{Shaded gray and dashed black:} limits on displaced decays of the scalar to electrons from the CHARM experiment. 
In shaded-gray we show limits with conservative assumptions regarding production rates and acceptances, while the region 
below the dashed-black line shows limits with aggressive assumptions. 
\textit{Yellow:} limits from Belle on $\br(B\to K\nu\bar{\nu})$.
\textit{Dashed-gray:} contours of constant scalar decay length, $c\tau_{\varphi}$.
}
\label{fig:higgsportal}
\end{figure}
 %%%%%%%%%%%%%%%%%%%%%%%%%%%%

In Fig. \ref{fig:higgsportal} we show in blue the contour of scalar mass $m_\varphi$ and mixing angle $\sin \theta$ for which the effective branching fraction Eq.~\eqref{eq:effectiveBR} is consistent with the central value of the KOTO measurement, Eq.~\eqref{eq:KOTOresult}.
In shades of blue we also show the regions of parameter space consistent with the measurement at $1\sigma$ and $2\sigma$.
\footnote{
The upper end of the $1\sigma$ band in Fig. \ref{fig:higgsportal} approximately coincides with the limits set by KOTO with previous datasets \cite{Ahn:2018mvc}.}
In dashed-gray lines we show contours of $c\tau_{\varphi}$.
The number of events measured at KOTO are consistent with a minimal Higgs portal model with mixing angles in the range $2 \cdot 10^{-4} \leq \theta \leq 10^{-3}$, across the mass range studied in this work.

There are a variety of constraints on the region of parameter space where the Higgs portal explanation is naively successful.  
The most obvious constraint comes from analogous decays $\br(K^+ \to \pi^+\,+\,\mathrm{inv.})$, 
which are normally related to the process of interest  at KOTO via the Grossman-Nir bound~\cite{Grossman:1997sk}.  
The NA62 and E949 experiments set constraints on these charged Kaon decays, 
which in the Higgs portal model arises from the diagram in Fig. \ref{fig:penguin}, 
with a width which can be obtained from Eq.~\eqref{eq:neutralkaon} with the replacement  $\mathrm{Re}\, g_{\varphi K \pi} \rightarrow |g_{\varphi  K \pi}|$ ~\cite{Leutwyler:1989xj, Gunion:1989we}.
NA62 sets a $95\%$ CL bound on the branching fraction ~\cite{NA62:2019}
%boe%
\begin{equation}
\label{eq:bound_na62}
\br(K^+ \to \pi^+ \nu \bar{\nu})_{\mathrm{NA62}}  <  2.44\times  10^{-10} \quad . \\
\end{equation}
%eoe%
In order to apply the NA62 limit we must take into account the effective branching fraction as done for KOTO in~Eq.~\eqref{eq:effectiveBR}.
The effective $\br^{\textrm{eff}}(K^+ \to \pi^\pm \nu \bar{\nu})$ measured at NA62 is given by Eq.~\eqref{eq:effectiveBR}, 
replacing neutral by charged mesons in the equation and accounting for the the experiment parameters.
For the NA62 detector size we use $L = 150\,\mathrm{m}$, 
while the scalar's energy is taken to be approximately half of the charged kaon energy at this experiment, 
$E_{\varphi} = 37\,\mathrm{GeV}$.
We neglect differences in efficiencies due to the different kinematics in the 3-body decay to neutrinos and the 2-body decay to our scalar, 
so for the NA62 effective branching fraction we set $\epsilon=1$.
The resulting limit is presented in Fig. \ref{fig:higgsportal} in red.
For scalar masses around the pion mass
there is a large gap in the bounds due to the large pion backgrounds, 
which are mitigated by kinematic cuts in the search, 
as anticipated for general BSM scenarios in ref.~\cite{Hou:2016den}.

Part of this gap is covered by a different NA62 analysis, which sets a limit on the invisible decays of the neutral pions from $K^+ \to \pi^+ \pi^{0}$ decays~\cite{NA62:2019}.
If our scalar has a mass very close to the pion mass, the process $K^+ \to \pi^+  \varphi$ mimics the invisible pion decay signature and is subject to this bound, modulo the $K \to \pi^+ \pi^0$ branching ratio of $20.6\%$.
We present the corresponding bound with a red column for scalar masses centered around the pion mass, with the width of the column set by the experimental pion energy resolution.

E949, on the other hand, reports $95\%$ CL bounds on the branching fraction of a charged Kaon into an invisible new particle $K^+ \to \pi^+ \varphi$, 
as a function of the particle's mass and lifetime \cite{Artamonov:2009sz},
so we can directly translate these bounds into the Higgs portal parameter space.
The bounds are shown in Fig. \ref{fig:higgsportal} in pink.
The bounds from NA62 and E949 rule out mixing angles $\theta \gtrsim \, 3 \cdot 10^{-4}$ for $m_\varphi \leq 200 \, \textrm{MeV}$ 
except in a gap of scalar masses around the pion mass, 
due to the aforementioned pion backgrounds.

Given the long-lifetime of the minimal Higgs portal, there are also potentially strong bounds from proton beam-dump experiments, the most relevant of which is the CHARM  experiment looking for displaced lepton decays from fixed target production at the CERN SPS~\cite{Bergsma:1985qz}.
At CHARM, 
our scalars are obtained from $B, K$ and $\eta$ meson decays, 
which are produced by the proton beam interactions on the fixed target. 
The event yield at the detector can be obtained using
\begin{multline}
N_{\mathrm{obs}} = 
\varepsilon_{\textrm{det}}  \, 
N_{\varphi}
\bigg(
e^{
-\frac{L_{\mathrm{dump}}}{c \tau_{\varphi}} 
\frac{m_{\varphi}}{p_{\varphi}} 
} 
-
e^{
-\frac{L_{\mathrm{dump}} + L_{\mathrm{fid}}}{c\tau_{\varphi}} 
\frac{m_{\varphi}}{p_{\varphi}} 
}
\bigg)
\quad ,
\label{eq:beamdump2}
\end{multline}
where $N_{\varphi}$ is the number of scalars falling within the CHARM angular (geometric) acceptance and $\varepsilon_{\textrm{det}}= 0.51 $ is the efficiency to detect the electron-positron pair.
The exponential factors in~\eqref{eq:beamdump2} determine the number of scalars that reach and decay within the detector volume.
$L_{\textrm{dump}}=480 \, \textrm{m}$ is the CHARM beam dump baseline, 
while $L_{\textrm{fid}}\,=\,35 \, {\textrm{m}}$ is the detector fiducial length. 
The scalar momentum is obtained assuming an average scalar energy of $E_{\varphi}=12.5 \, \textrm{GeV}$.
This energy is obtained by assuming that the scalar takes half the energy of the parent meson, 
and that the parent meson's energy is similar to the typical $25 \, \textrm{GeV}$ pion energy reported in \cite{Bergsma:1985qz}.
The number of scalars falling within the detector's solid angle $N_{\varphi}$ has uncertainties 
inherited from the uncertainties of the parent meson rates, momentum and angular distributions.   We have found that these uncertainties are often underappreciated and in an appendix we provide details about the underlying assumptions for hadron beam dumps applicable to all BSM models which share production mechanisms from meson decays.  In order to provide a realistic interpretation of the CHARM limits,
we present two bounds, 
one with conservative and one with aggressive assumptions.

The resulting CHARM bounds are presented in shaded gray in Fig. \ref{fig:higgsportal} for our conservative assumptions,
and as a black-dashed contour for our aggressive assumptions.
The conservative bounds rule out $\theta \gtrsim  2 \cdot 10^{-3}$ across our range of masses. 
Note that CHARM bounds cover a large fraction of the pion mass gap window left by NA62 and E949.

We now comment on sub-leading bounds on the Higgs portal model relevant for our range of masses. 
First, Belle sets bounds on the $B \to K \nu\bar{\nu}$ decay ~\cite{Grygier:2017tzo}, 
%boe%
\begin{equation}
\label{eq:bellelimit}
\br(B\to K\nu\bar{\nu}) < 1.6\times 10^{-5} \quad .
\end{equation}
%eoe%
In the minimal Higgs portal, 
this decay arises at one loop via up-type quark mediated penguin diagrams, 
with the scalar escaping the Belle undetected. 
To minimize the hadronic uncertainties, 
we compute this branching ratio by considering the ratio~\cite{Chivukula:1988gp, Grinstein:1988yu}
%boe%
\begin{multline}
\frac{\Gamma(B \to K \varphi)}{\Gamma(B \to X_c e \nu)} = \frac{27}{64\pi^2 m_b^2} \frac{1}{f_{c/b}} \left( 1 - \frac{m_s^2}{m_b^2}\right)
\\ 
\times 
\bigg |
\sum_{f = u, c, t}  \frac{ V_{fs}^* V^{\phantom{*}}_{fb} }{V_{cb}}  \lambda_{\varphi}^f m_f \,  
\bigg|^2,
\end{multline}
%eoe%
where the scalar couplings $\lambda_{\varphi}^f$ are given in Eq.~\eqref{eq:higgsportalcouplings}, and $f_{c/b}=0.51$.
We then normalize this result to the experimentally measured ratio $\br(B\to X_c e\nu) = 0.104$ for both $B^0$ and $B^{\pm}$~\cite{Patrignani:2016xqp}.
We compare our branching fraction calculation with the limit in Eq.~\eqref{eq:bellelimit}, and present the bound in Fig. \ref{fig:higgsportal} in yellow.

Second, the KTeV collaboration sets a limit on the branching fraction of Kaons to a pion and electron-positron pairs \cite{AlaviHarati:2003mr},
\begin{equation} 
\label{eq:KTeVelectrons}
\br(K_L \to \pi^0 e^+ e^-) \leq 2.8 \cdot 10^{-10} \quad ,
\end{equation} 
which in our model is generated from $\br(K_L \to \pi^0 \varphi)$ followed by scalar decays. 
While the minimal Higgs portal scalar does decay mostly into electrons, 
in the range of mixing angles allowed by the CHARM and charged kaon factory bounds,
it is rather long-lived. 
As a consequence, 
most scalars produced from Kaon decays escape the KTeV fiducial volume unobserved. 
To confirm this expectation,
we perform an \textit{aggressive} estimate of the number of two-electron events effectively observed at KTeV, 
by assuming that Kaons decay at rest
\footnote{The actual KTeV analysis requires the Kaons to have a significant boost, which only weakens this bound.} at the beginning of the $L_{\textrm{KTeV}} = 60 \, \textrm{m}$ detector length into our scalars.
We then apply an exponential decay factor to obtain the number of scalars that decay into electrons within the detector.
For mixing angle values $\theta = 10^{-3}$, which are already excluded by CHARM and NA62/E949, 
we find that the branching fraction $\br(K_L \to \pi^0 e^+ e^-)$ effectively observed at KTeV due to scalar decays is an order of magnitude below the reported bound Eq.~\eqref{eq:KTeVelectrons}.
We conclude that this bound is sub-leading.

Finally, 
for low scalar masses around an MeV, 
bounds from BBN apply. 
However, these bounds depend on the value of the scalar-Higgs quartic coupling, 
and on assumptions regarding the reheating temperature \cite{Fradette:2017sdd}, 
so they are not presented here. 

By comparing the regions of Higgs portal parameter space favored by the KOTO measurement and the leading beam dump and flavor bounds presented in Fig.\ref{fig:higgsportal},
we conclude that the Higgs portal may explain the central value of the KOTO anomaly in a region of parameter space around $m_\varphi \simeq 120 \, \textrm{MeV}, \theta \simeq 5\cdot 10^{-4}$. 
More broadly, including the $2\sigma$ bands of the KOTO measurement, we find that the Higgs portal leads to a realistic explanation of the KOTO results even above or below the pion mass gap.

 %%%%%%%%%%%%%%%%%%%%%%%%%%%%
 %%%%%%%%%%%%%%%%%%%%%%%%%%%%
\section{Hadrophilic scalars and the KOTO excess}

 %%%%%%%%%%%%%%%%%%%%%%%%%%%%
\begin{figure}[h!]
\includegraphics[width=8cm]{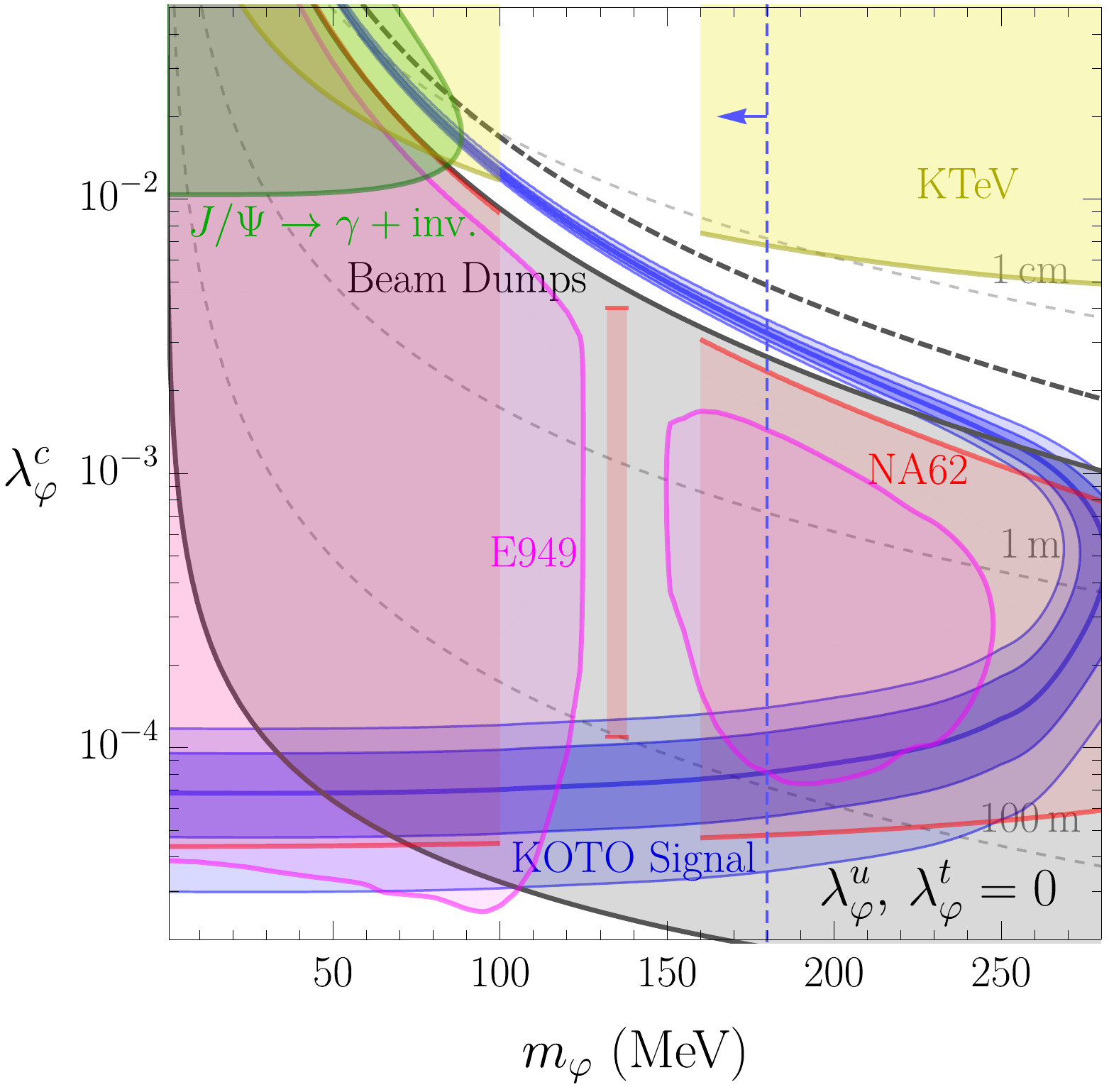}
\caption{
Charm-philic scalar interpretation of the KOTO excess and leading bounds on the model, 
plotted as a function of the scalar's mass and its charm-Yukawa coupling.
The remaining scalar-quark and lepton Yukawas have been set to zero.
\textit{Blue:} regions of parameter space consistent with the number of $K_L \to \pi^0 \nu \bar{\nu}$ events observed at KOTO.
The solid line corresponds to the measured central value, while the shaded regions include the $1$ and $2\sigma$ compatible values. 
The region to the left of the vertical-dashed blue line corresponds to singlet masses which are consistent with the kinematics of the observed KOTO events.
\textit{Red:} 
limits from NA62 on $\br(K^+ \to \pi^+  \nu \bar{\nu})$, 
and $\br(K^+ \to \pi^+  \pi^0)$ with $\pi^0$ decaying invisibly.
\textit{Pink:} 
limits from E949 on $\br(K^+ \to \pi^+  \nu \bar{\nu})$.
\textit{Shaded gray and dashed black:} limits on displaced decays of the scalar to photons from the NuCal beam-dump experiment. 
In the gray shaded region we present a conservative estimate of the NuCal bounds,
while the regions below the dashed-black line may be excluded with more aggressive assumptions regarding scalar production at the fixed target.
\textit{Green:} limits from the Crystal Ball detector on $\br(J/\Psi \to \varphi \gamma)$. \textit{Yellow:} limits from the KTeV experiment from  $\br(K_L \to \pi^0 \gamma\gamma)$.
\textit{Dashed gray:} decay length $c \tau_{\varphi}$ of the singlet-like scalar.
}
\label{fig:region_plot}
\end{figure}
%%%%%%%%%%%%%%%%%%%%%%%%%%%%
%%%%%%%%%%%%%%%%%%%%%%%%%%%%%%%%%%%%%%%%%%%%%%%%%%%%%%%% 

To analyze scalar-singlet extensions of the Standard Model within a more general flavor framework, 
we consider the hadrophilic model presented earlier, with non-renormalizable interactions, Eq.~\eqref{eq:dim5_yukawas}.
This allows us to study novel hierarchies in the couplings of the scalar singlet to the different SM quark generations.
To simplify the phenomenology, 
we set the Higgs-scalar mixing angle to zero, 
$\theta=0$, 
so that the Yukawa couplings of our singlet to quarks are exclusively given by Eq.~\eqref{eq:hadrophilicYukawas}.
In this case, 
our scalar does not couple to down-type quarks nor leptons, 
so for the range of masses that we now explore it may only decay to two photons via one-loop up-type quark mediated diagrams. 
The corresponding decay rate is given by 
%boe%
\begin{equation}
\Gamma_{\varphi \to \gamma\gamma} = \frac{3 \alpha^2 m_\varphi^3}{256\pi^3} \sum_{q=u,c,t} \, \bigg[\frac{\lambda_{\varphi}^q | F_{1/2} |}{m_q} \bigg]^2 \quad ,
\label{eq:photons}
\end{equation}
%eoe%
where $\alpha$ is the fine-structure constant, 
$\lambda_{\varphi}^q$ are the up-type quark Yukawa couplings of our hadrophilic scalar defined in Eq.~\eqref{eq:hadrophilicYukawas},
and $F_{1/2}$ is the usual fermionic loop function familiar from Higgs decays~\cite{Gunion:1989we}.

We first explore a hadrophilic scalar coupling exclusively to the charm quark, 
as this has a very distinctive phenomenology compared to the minimal Higgs portal.  
For the charm-phillic scalar we
set $\lambda_{\varphi}^u=\lambda_{\varphi}^t=0$ in Eq.~\eqref{eq:hadrophilicYukawas}.
In this scenario, 
our scalar is produced at KOTO from the penguin diagram Fig. \ref{fig:penguin} with an internal charm quark.
We calculate the number of events from this process effectively tagged as $K_L \to \pi^0 \nu \bar{\nu}$ in KOTO as in the previous section, 
and in Fig. \ref{fig:region_plot} in blue, 
we show the regions of parameter space consistent with the KOTO measurement. 
We present the results as a function of the scalar-charm Yukawa $\lambda_{\varphi}^c$ and its mass.
In the figure we also show contours of $c\tau_{\varphi}$ obtained from Eq.~\eqref{eq:photons} in dashed-gray.
We identify two ranges of values for the scalar-charm Yukawa that can accommodate the KOTO anomaly.
First, we find a band of sizable Yukawas, $\lambda_{\varphi}^c \geq 10^{-3}$, 
where the scalar production rate from $K_L$ decays is large, 
but the number of events reconstructed as $K_L \to \pi^0 \nu \bar{\nu}$ at KOTO is exponentially suppressed since the scalar decays into photons before reaching the detector.
Second, we find a band where Yukawas are small, $3 \cdot 10^{-5} \leq \lambda_{\varphi}^c \leq 10^{-4}$.
For this range of Yukawas the scalar production rate is small, but the scalar lifetime is large, 
so most scalars escape the KOTO fiducial volume unobserved 
and are thus tagged as $K_L \to \pi^0 \nu \bar{\nu}$.
Note that for the values of scalar-charm Yukawa couplings that can explain the KOTO excess, 
the scalar's lifetime is comparatively much shorter than the lifetime of the minimal Higgs portal scalar studied in the previous section. 
This leads to significant differences in the bounds that apply to the charm-philic and the minimal Higgs portal scalars, as we will now see.

In Fig. \ref{fig:region_plot} we present the leading bounds on the charm-philic model. 
Bounds from NA62 and E949 from the limits on $\br(K^+ \to \pi^+\,+\,\mathrm{inv.})$  are calculated as in the previous section and presented in red and pink.
We observe that for large $\lambda_{\varphi}^c$ these bounds disappear,
since the scalar decays into photons before reaching the corresponding detectors.  This is a specific model-realization of the finite lifetime effects discussed in \cite{Kitahara:2019lws}, 
that we refer to as a ``lifetime gap''.

Additional bounds on the charm-philic scenario are set by the Crystal Ball detector, 
which sets a limit \cite{Edwards:1982zn}
\begin{equation}
\label{eq:crystalball}
\br(J/\psi \to \varphi \gamma) < 1.4\times 10^{-5} \quad ,
\end{equation}
where $\varphi$ escapes the detector invisibly. 
We compute the branching fraction $\br(J/\psi \to \gamma \varphi)$ using the standard relation~\cite{Wilczek:1977zn}:
%boe%
\begin{equation}
\Gamma(J/\psi \to \gamma \varphi) = \Gamma(J\psi \to \mu^+\mu^-)\, \frac{(\lambda_{\varphi}^c)^2}{2\pi\alpha}\,C_{J/\psi} \quad ,
\label{eq:Jpsiwidth}
\end{equation}
%eoe% 
where $C_{J/\psi}$ is a factor that encodes both QCD and relativistic corrections~\cite{Nason:1986tr} and $\lambda_{\varphi}^c$ is the scalar-charm Yukawa. 
We consevatively take $C_{J/\psi} = 0.1$, 
a value that is in agreement with the first order corrections in ref.~\cite{Nason:1986tr}. 
We apply the bound~\eqref{eq:crystalball} by using~\eqref{eq:Jpsiwidth}, $\Gamma_{J/\psi}^{\textrm{SM}}=92.9 \, \textrm{keV}$ \cite{Patrignani:2016xqp},
and applying and exponential suppression factor to account for the scalars decaying to photons before escaping the $25 \, \textrm{cm}$ radius Crystal Ball.
The resulting limit is shown in Fig. \ref{fig:region_plot} in green.

Given the lifetime of charm-philic scalar, hadronic beam-dump experiments can also set stringent bounds.  However, the CHARM experiment, which was the most relevant bound for the minimal Higgs portal,
is impuissant in this scenario due to the scalar's short lifetime compared to the experiment's long baseline.
The strongest beam-dump bounds instead come from NuCal~\cite{Blumlein:1990ay, Blumlein:2011mv, Blumlein:2013cua}, 
which has a shorter baseline than CHARM and lower beam energy.
The bounds are obtained as for CHARM in the previous section, using Eq.~\eqref{eq:beamdump2} with
with $N_{\textrm{POT}} = 4\times 10^{17}$,
beam-dump baseline $L_{\textrm{dump}}=64 \, \textrm{m}$ and fiducial length $L_{\textrm{fid}}=23 \, \textrm{m}$.
Again, to account for the uncertainties in the scalar production rate we present both a conservative and an aggressive bound, 
with different assumptions on scalar production at NuCal, 
which we discuss in the appendix.
In Fig. \ref{fig:region_plot}, we present the conservative bound in shaded gray and the aggressive bound with a dashed-black contour.

Finally, we comment on sub-leading bounds on the charm-philic scenario. 
MAMI sets a constraint on $\br(\eta \to \varphi \pi^0) < 3\times 10^{-4}$ ~\cite{McNicoll:2010qk, Nefkens:2014zlt}. 
We calculate the corresponding bound in our model by using the chiral lagrangian approximation to obtain the scalar-$\eta$ coupling as detailed in the appendix,
and find that  it is weaker than the bounds discussed above.
The same conclusion applies to constraints on $B \to K \nu\bar{\nu}$ from Belle~\cite{Grygier:2017tzo}. 
Finally, KTeV has measured $\br(K_L \to \pi^0\gamma\gamma) = (1.29 \pm 0.06) \times 10^{-6}$~\cite{Abouzaid:2008xm}.
In our model, the same final state is obtained from $K \to \pi^0\varphi$ with $\varphi \to \gamma \gamma$.  While this bound wasn't relevant for the minimal Higgs portal, for the shorter lifetimes in this scenario it can potentially apply.
Unfortunately, it is not possible to directly apply the KTeV measurement of $\br(K \to \pi^0 \gamma\gamma)$ as a direct bound on $\br(K \to \pi^0\varphi)$ in our model, 
as the measurement assumes that the two photons and the pion originate at the same vertex, 
while our scalar decays displaced due to the large boost inherited from the parent $K_L$ meson.
In the absence of a detector simulation, 
we obtain a conservative bound by considering only the scalar decays that appear prompt given KTeV's vertex resolution, 
which we take to be $25\,\textrm{cm}$, based on the bin-widths for decay locations given in~\cite{Abouzaid:2008xm}.
We further assume that the scalars have an average energy of $50$ GeV.  
The resulting bound is shown in yellow in Fig. \ref{fig:region_plot}.

From Fig. \ref{fig:region_plot}, we see that the KOTO result may be explained by a charm-philic scalar with masses in the range $100 \, \textrm{MeV} \leq m_{\varphi} \leq 180 \, \textrm{MeV}$ and Yukawas in the range $3 \cdot 10^{-3} \leq \lambda_{\varphi}^c \leq 10^{-2}$.
For these range of parameters, the observation of $K_L \to \pi \nu \bar{\nu}$ events at KOTO is consistent with strong bounds from charged kaon factories due to the aforementioned ``lifetime gap".
However, we find that in the charm-philic scenario,
the lifetime gaps consistent with the KOTO excess may be completely covered by NuCal bounds.
In fact, 
while the conservative estimate of the NuCal bounds in shaded gray is consistent with KOTO, 
the aggressive estimate in dashed-black completely rules out the explanation. 
In order to determine which bound is the most realistic, 
a dedicated study of tails of meson momentum distributions at fixed target experiments is needed, as described in the appendix,
which is beyond the scope of this work.

We conclude by commenting on up-philic and top-philic scalars. 
In the up-philic case, $\lambda_{\varphi}^c = \lambda_{\varphi}^t = 0$,
the penguin diagram Fig. \ref{fig:penguin} leading to $K_L \to \pi \nu \bar{\nu}$ is mediated by internal up-quark loops and is strongly suppressed by one up-quark mass insertion.
While it is possible to explain the number of events observed at KOTO in this scenario, 
doing so requires a large up-quark scalar Yukawa, which is excluded by various experiments~\cite{Batell:2018fqo}.
In the top-philic case, $\lambda_{\varphi}^u=\lambda_{\varphi}^c=0$, 
the situation is similar to the minimal Higgs portal setup presented in Fig. \ref{fig:higgsportal}, 
with $\sin\theta$ replaced by $\lambda_{\varphi}^t$.
In this situation, 
the KOTO events are again consistent with bounds from charged kaon factories mostly in a region of masses around the pion mass.
Neither the up-philic nor top-philic scenarios lead to any additional regions of parameter space consistent with the KOTO excess due to the ``lifetime gap" suggested in~\cite{Kitahara:2019lws}. Thus, the only ``lifetime gap" present for a hadrophilic scalar model is in the charm-philic scenario, which is in tension with current bounds from NuCal.
Allowing for different combinations of $\lambda_{\varphi}^u, \lambda_{\varphi}^c$ and $\lambda_{\varphi}^t$ simultaneously nonzero does not modify this conclusion.

%%%%%%%%%%%%%%%%%%%%%%%%%%%%%%%%%%%%%%%%%%%%%%%%%%%%%%%%
%%%%%%%%%%%%%%%%%%%%%%%%%%%%%%%%%%%%%%%%%%%%%%%%%%%%%%%%
\section{Conclusions}

In this work we investigated a possible new physics explanation of the observed KOTO excess in the process $K_L \to \pi \nu \bar{\nu}$. 
In our setup, the neutrinos are replaced by a singlet scalar that escapes the detector invisibly, and 
we find that simple models with light new scalars may account for the excess.
Interestingly, the simplest possible extension of the Standard Model, 
the minimal Higgs portal, can explain the anomaly. 
Models with hadrophilic scalars were also studied, 
and we found that a top or charm-philic scalar could also be the origin of the excess.

If the observed events are due to new physics, a similar number of events should be observed in future KOTO datasets. 
If the scalar lies outside the $100 \lesssim m_{\varphi} \lesssim 160\,\mathrm{MeV}$ vetoed window, it should also be visible in future $K^+ \to \pi^+\,+\,\mathrm{inv.}$ searches at NA62 as well. 
Our results demonstrate that extending the searches from $\pi \to \mathrm{inv.}$ to cover the entire vetoed region would be extremely useful for constraining light new physics solutions of the excess.
We found that hadronic beam-dump experiments may also be efficient at testing these solutions, 
but suffer from uncertainties in the production rates and acceptances, discussed in detail in the appendices.
For this reason, 
it is especially interesting to consider lifetime frontier experiments where the production mechanism is under better theoretical control. 
For example, the MATHUSLA experiment only relies upon the knowledge of the hard QCD production process~\cite{Curtin:2018mvb}, and amusingly is also the {\em most} sensitive to the minimal Higgs portal in the parameter space which explains the KOTO excess~\cite{Beacham:2019nyx}.
In addition, 
we found that Higgs portal and top-philic model explanations of the anomaly share similarities and are challenging to distinguish at KOTO, 
but they could be distinguished by future long-baseline experiments by probing differences in the corresponding di-electron or di-photon final states.

Models with light new scalars can be accommodated in well-motivated UV constructions. 
Naively, a real scalar in the sub-GeV range appears tremendously tuned.
However, if for instance a large extra dimensional scenario is invoked to solve the Higgs hierarchy problem \cite{ArkaniHamed:1998rs}, 
it automatically mitigates the hierarchy problem of the new scalar.
In addition, a new scalar with similar couplings to ones described for the minimal Higgs portal
could correspond to the Goldstone of broken scale symmetry \cite{Goldberger:2008zz,Coradeschi:2013gda}, 
or perhaps be the radion responsible for stabilizing extra dimensions \cite{Rattazzi:2000hs,Bellazzini:2012vz}.  
Naively, a sub-GeV scale dilaton portal is hard to achieve without tuning \cite{Bellazzini:2012vz}, 
but is certainly worth exploring further if the KOTO excess persists. 
Within supersymmetry, complex singlet fields are accommodated in the NMSSM.
Supersymmetry breaking mass-splitting may be then introduced to keep only the real scalar at the sub-GeV scale while decoupling the pseudoscalar singlet component,
or if the pseudoscalar and singlino components are also close to the MeV scale, the corresponding phenomenology could also be of interest.
Hadrophilic light scalars consistent with strong bounds from FCNCs, on the other hand, may arise in flavor-aligned UV completions with an extra Higgs doublet or new vector-like quarks,
where flavor alignment is ensured by an SFV flavor construction, which also solves the strong-CP problem \cite{Egana-Ugrinovic:2018znw}.

The experimental result obtained by KOTO may ultimately be due to statistics or unaccounted backgrounds. 
Nevertheless, 
we have demonstrated that from a purely theoretical perspective the observation is incredibly simple to explain,
and is motivated by interesting UV constructions.

\section{Acknowledgments}
We would like to thank Kohsaku Tobioka for helpful comments on the manuscript. 
We would also like to thank Asimina Arvanitaki, Yang-Ting Chen, Rouven Essig, Evgueni Goudzovski, Junwu Huang, George Sterman and Michael Wilking for helpful discussions.

DEU is supported by Perimeter Institute for Theoretical Physics. 
Research at Perimeter Institute is supported in part by the Government of
Canada through the Department of Innovation, Science
and Economic Development Canada and by the Province
of Ontario through the Ministry of Economic Development, Job Creation and Trade. 
DEU thanks the Galileo Galilei Institute for Theoretical Physics for the hospitality and the INFN for partial support during the completion of this work.  The work of SH and PM was supported in part by the National Science Foundation grant PHY-1915093.  PM would like to thank the Mt. Sinai West maternity ward for its hospitality in the final stages of this project.

%%%%%%%%%%%%%%%%%%%%%%%%
\newpage
\section{Appendix}

\subsection{Production of light scalars at hadronic beam dump experiments}

In this appendix we review the production of light bosons in hadronic beam dump experiments arising from meson decays.  While there are other possible production mechanisms such as Bremsstrahlung and hard production, these only exacerbate the difficulties we present below and increase any putative bounds.  For the range of masses of interest for our scalars, meson decays are the most robust bound for beam dumps as exemplified by the numerous studies in this channel.  Nevertheless even for this ``robust" bound, there are a number of uncertainties that we wish to emphasize as they can change the bounds significantly.  
We will focus on the production of a light scalar, but most of the discussion below applies for any light new particle that can be produced in meson decays, such as dark photons or axion-like particles, and it would be useful to extend this discussion further in the context of specific models. 
In the following we focus on the parameters for the CHARM and NuCal proton beam-dump experiments which are most relevant for the models considered, but the generic lessons apply to other existing and proposed hadronic beam dump experiments.

The number of light scalars that are produced from meson decays at a given hadronic beam dump experiment is given by~\cite{Barabash:1992uq, Winkler:2018qyg}
\begin{eqnarray}
\label{eq:beamdumpprod}
\nonumber
N_{\varphi} & = & 
N_{\textrm{POT}}\, 
\,
\big[
\varepsilon_{\textrm{geom}}^K
n_K\,D_K\,
\br({K\rightarrow \pi \varphi})
\, \\
 \nonumber
&+&
\varepsilon_{\textrm{geom}}^\eta
n_\eta
 \br({\eta \rightarrow \pi \varphi})  
 +
 \varepsilon_{\textrm{geom}}^B
 n_B
 \br({B \rightarrow \pi \varphi})
 \big]\\
 D_K & = & \ell_K / ( \gamma_K c \tau_K )
\end{eqnarray}
where $N_{\textrm{POT}}$ is the number of protons on target, $n_{K,\eta,B}$ are the average number of mesons produced per POT, $ \gamma_K $  is the average Kaon boost, and $\varepsilon_{\textrm{geom}}^{K,\eta,B}$ are probabilities (acceptances) for the scalar to fall within the detector solid angle, given that it originated from a $K, \eta$ or $B$ meson decay.
$D_K$ encodes the suppression in the scalar production rate due to Kaons being reabsorbed in the target before decaying, and 
$\ell_K$ is the Kaon absorption length, which depends on the target material
\footnote{Eq.~\eqref{eq:beamdumpprod} is valid in the limit where $\ell_H$ is taken to be much bigger than the CHARM target thickness.}.
For CHARM we take $\ell_K = 15.3\,\textrm{cm}$~\cite{Patrignani:2016xqp}, while for NuCal we use $\ell_K = 27.4\,\textrm{cm}$~\cite{Barabash:1992uq}.
The branching ratios for $K$ and $B$ decays can be computed as described in the text, while 
$ \br({\eta \rightarrow \pi \varphi})$ is calculated using chiral perturbation theory, matching the scalar-gluon and scalar-quark couplings to the low energy chiral Lagrangian, 
as detailed in the following appendix.
Note that we neglect sub-leading CP violating production from the decay of $K_S$.

While the branching ratios can be computed, the meson multiplicities and momentum distributions (required to compute the geometric acceptance) for the different mesons must be taken from data, and can change substantially at different energies.   This is the analogous problem for neutrino beam experiments, where if the underlying meson distributions were better known, the neutrino beams could be better characterized. Unfortunately, the SM inputs suffer from large uncertainties, often in the realm most relevant for setting limits on long-lived BSM particles.

Charged pion and Kaon multiplicities have been measured at the SPS in refs.~\cite{AguilarBenitez:1991yy}, 
but a large range of values have been cited in the literature. 
For the $\pi^+$ multiplicity, for example, values in the range 
$1.86 - 3.6$ 
have been used~\cite{Ammosov:1976zk, Barabash:1992uq, Blumlein:2011mv} at the $70\,\mathrm{GeV}$ energies relevant for NuCal. 
The $\pi^0$ and $\eta$ multiplicities have also been measured in ref.~\cite{AguilarBenitez:1991yy}, but the neutral Kaon multiplicities (and the $\pi^0$ multiplicity at other energies) are usually assumed to be the average of the positive and negatively charged values. 
For the purposes of this work, we have taken the multiplicities at CHARM to be $n_K = 0.33, 0.22 \textrm{ and } 0.28$ 
for the $K^+$, $K^-$ and $K_L$ respectively, and $n_{\eta} = 0.31$ based on ref.~\cite{AguilarBenitez:1991yy}.
For NuCal, the Kaon multiplicities are taken directly from ref.~\cite{Barabash:1992uq} to be $n_{K^\pm}=0.5$, $n_{K_L}=0.3$.
The $\eta$ multiplicity is obtained by noting that the $\eta$ to $\pi$ multiplicity ratio is roughly independent of the center of mass energy \cite{Albrecht:1995ug,Adler:2006bv}, 
and is given by $n_{\eta}/n_{\pi}=0.078$~\cite{AguilarBenitez:1991yy}, with $n_{\pi}=2.35$  at NuCal energies \cite{Ammosov:1976zk}. 
This gives a multiplicity $n_{\eta}=0.18$ at NuCal.

For the bounds set in this paper, we do not include production from $B$ meson decays, 
due to the large uncertainties in the inclusive $b\bar{b}$ production cross section measurements~\cite{Lourenco:2006vw}, 
from which the $B$ meson multiplicities are obtained.
We have checked that taking a multiplicity of $n_B = 10^{-7}$, consistent with the assumptions in ref.~\cite{Anelli:2015pba}, with a geometric acceptance similar to the acceptance used for Kaons only modifies our bounds at the percent level.

 %%%%%%%%%%%%%%%%%%%%%%%%%%%%
\begin{figure*}
\includegraphics[width=0.48\linewidth]{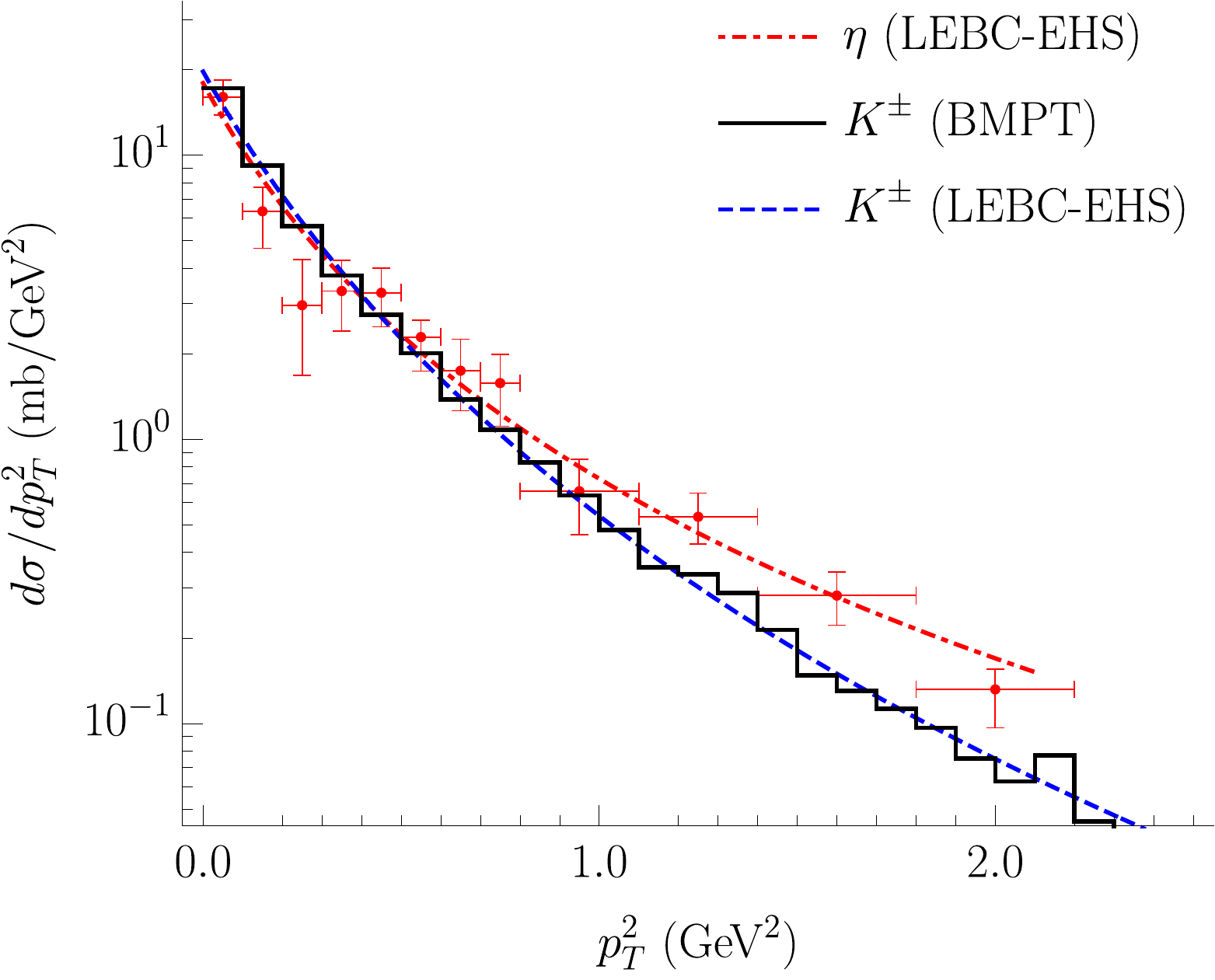}
~~~
\includegraphics[width=0.48\linewidth]{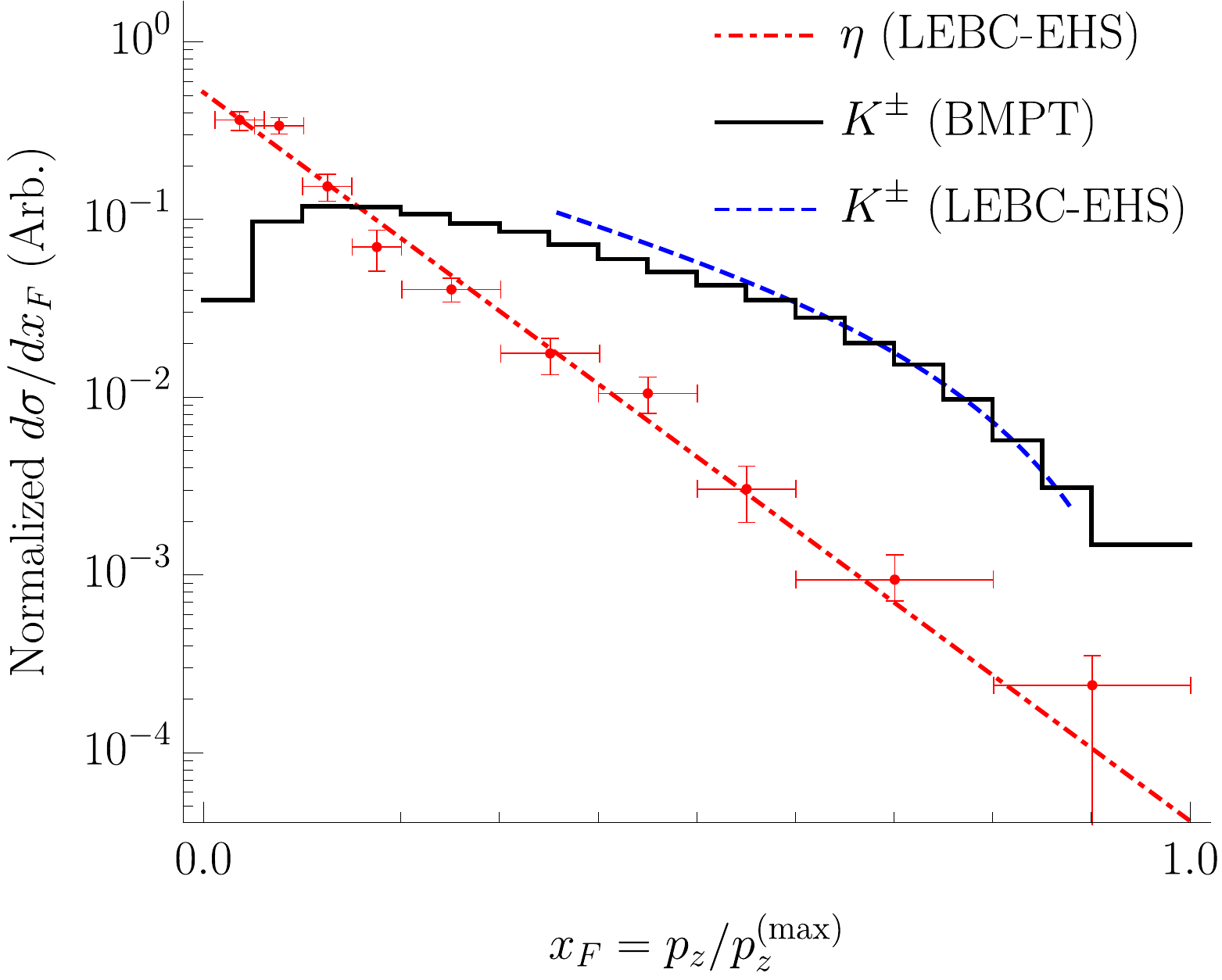}
\caption{Comparison of the $p_{T}^2$ (left) and $x_F$ (right) distributions of $\eta$ and $K$ mesons produced in $400\,\textrm{GeV}$ $pp$ interactions. 
The red data points and dot-dashed curve show the measured $\eta$ distribution from the LEBC-EHS collaboration at the CERN SPS~\cite{AguilarBenitez:1991yy} along with their best fit. 
The black stepped histogram shows the distribution obtained using our simulations based on the \texttt{BdNMC} code~\cite{deNiverville:2016rqh} using the BMPT distributions~\cite{Bonesini:2001iz}.
The dashed blue curve shows the average of the fit to $K^+$ and $K^-$ data from the LEBC-EHS results, which were fit over the full range of $p_T^2$, but only for $x_F \in [0.36, 0.88]$.
}\label{fig:distributions}
\end{figure*}

To compute the geometric acceptance at the detector, as well as the survival probability as a function of the scalar lifetime, the momentum distribution of the scalars must also be estimated.
Due to the large boost of the mesons at CHARM and NuCal, 
the scalar is produced approximately along the direction of the parent meson, and we take the scalar energy to be half the parent meson energy to obtain the boost. 
The geometric acceptances, particularly for the off-axis CHARM detector, can be quite sensitive to the transverse momentum distribution.
Furthermore, for scalars with shorter lifetimes in the $\textrm{cm} - \textrm{m}$ range (such as the charm-philic scalar with  couplings $\lambda_{\varphi}^c \sim 10^{-2}-10^{-3}$), the survival probabilities at a distant detector are particularly sensitive to the momentum distributions.
This is because these lifetimes are short compared to the baselines at NuCal ($L = 64\,\textrm{m}$) and CHARM ($L = 480\,\textrm{m}$), so only highly boosted scalars coming from energetic parent mesons make it to the detector, and the bounds are exponentially sensitive to the tail of the meson momentum distributions.
All of these distributions must be taken from data, and suffer from large uncertainties particularly for the neutral mesons and at high longitudinal momentum.

At CHARM energies, the double-differential pion and Kaon momentum distributions have been measured and parameterized in ref.~\cite{Bonesini:2001iz}, and these distributions can be implemented in the \texttt{BdNMC} code~\cite{deNiverville:2016rqh}.
We use this code to estimate the geometric acceptance for Kaon initiated production, and find $\varepsilon_{\textrm{geom}}^K=3\cdot 10^{-2}$.
The Kaon momentum distribution at CHARM is seen to be quite similar to the pion distributions~\cite{AguilarBenitez:1991yy},
so we checked that our acceptance is consistent with the pion acceptance reported in ref.~\cite{Bergsma:1985qz}, assuming the $\pi^0$ multiplicity measured at the SPS.
This acceptance is an order of magnitude larger than the one obtained in ref.~\cite{Winkler:2018qyg}, which we believe to be a result of differences between Pythia and data in the forward direction (see also ref.~\cite{Dobrich:2019dxc}).
For the $\eta$ momentum distribution, 
to the best of our knowledge, only single differential distributions have been reported \cite{AguilarBenitez:1991yy}, 
while double-differential distributions are required to reliably calculate the geometric acceptance. 
As shown in Fig.~\ref{fig:distributions}, the $\eta$ distributions tend to have a larger number of events at high transverse momentum as compared to Kaons \cite{AguilarBenitez:1991yy}, 
so we expect a smaller geometric acceptance for scalars coming from $\eta$ decays at CHARM, relative to scalars produced from Kaons.
The results of ref.~\cite{AguilarBenitez:1991yy} also demonstrate that the shape of the longitudinal momentum distributions can be quite different for Kaons and $\eta$'s, which has important implications for the survival probabilities we discuss below.
Given the lack of data, 
an aggressive approach is to assume
 the same acceptance for scalars produced from $\eta$'s and Kaons, 
$\varepsilon_{\textrm{geom}}^\eta = \varepsilon_{\textrm{geom}}^K=3\cdot 10^{-2}$.
For NuCal, on the other hand, we take a geometric acceptance of $19\%$, based on the pion acceptance in ref.~\cite{Blumlein:2011mv}. 

To compute the survival probabilities, a conservative approach is to assume that the mesons have a ``typical energy" that agrees with the experimentally reported bounds, and compute the number of scalar decays within the detector's fiducial volume using Eq.~\ref{eq:beamdump2}, using a detector size of $23\,\textrm{m}$ for NuCal and $35\,\textrm{m}$ for CHARM.
This is a conservative scenario, because it disregards that a significant fraction of the mesons will have much higher momentum, and thus a much larger survival probability at the distant detector. 
However, we have verified that using a typical energy of $18\,\textrm{GeV}$ for the mesons produced at the beam reproduces the experimental bounds on a light, SM-like Higgs presented in ref.~\cite{Barabash:1992uq} using NuCal data.

A more aggressive approach is to estimate the number of scalars decaying at the detector by integrating over the full momentum distribution.
This distribution was estimated for NuCal in ref.~\cite{Blumlein:2011mv}, extrapolating the measured pion distributions in~\cite{Ammosov:1976zk}, and we assume the Kaon momentum distribution to be similar.
This extrapolation was compared to a simulated distribution using \texttt{BdNMC} with the BMPT distribution, and while good agreement was found for $p_L(\pi) \lesssim 35\,\mathrm{GeV}$, the extrapolation estimated a larger number of pions for larger momenta. 
Due to the longer baseline as compared with NuCal, CHARM only sets limits for scalars with a much longer lifetime, for which the survival probability becomes much less sensitive to the exact shape of the distribution.

In light of all the uncertainties discussed above,
 for each of the two models discussed in this work we consider both an ``aggressive" and a ``conservative" limit.
For the Higgs portal, the leading constraints come from CHARM, due to the longer lifetime.
For the conservative limit, we neglect the production from $\eta$ decays entirely, due to the uncertainties both in the distribution and on the branching ratio $\br({\eta \rightarrow \pi \varphi})$ discussed in the following appendix, and consider scalar production only from Kaon decays.
We use $N_{\textrm{POT}} = 2.4\times 10^{18}$ and the geometric acceptance and multiplicities above to compute the expected number of scalars as a function of mass and mixing angle using Eqs.~\ref{eq:beamdump2} and \ref{eq:beamdumpprod}, using a typical energy of $25\,\mathrm{GeV}$ for the Kaons, based on the pion distribution in ref.~\cite{Bergsma:1985qz}\footnote{
Using either the typical energy of $25\,\mathrm{GeV}$ or the BMPT distribution~\cite{Bonesini:2001iz} was verified to reproduce the bounds on axion like particles set by the CHARM experiment in~\cite{Bergsma:1985qz}.}.
In the aggressive case, we also include the production from $\eta$ decays, with the assumptions described above and the branching ratio computed in the following appendix. 
The conservative and aggressive bounds are respectively shown as the gray shaded region (bounded by the solid black curve) and dashed black curve in Fig.~\ref{fig:higgsportal}.

For the charm-philic scenario, the leading constraints come from NuCal, with $N_{\textrm{POT}} = 1.7\times 10^{18}$.
In the conservative scenario, we again consider only Kaon decays, and assume all parent mesons to have the typical energy of $18\,\mathrm{GeV}$ as described above.
In the aggressive approach, we take the pion momentum distribution in ref.~\cite{Blumlein:2011mv} for all the parent mesons, and include both Kaon and $\eta$ decays, integrating over the meson energy to compute the survival probability in Eq.~\ref{eq:beamdump2}.
The corresponding conservative and aggressive bounds are given in Fig.~\ref{fig:region_plot} by the shaded black region, and dashed-black contour respectively. 
We see that the constraints in the conservative and aggressive scenario are quite different, and have important implications for interpretations of current results and projected constraints from future experiments.

%%%%%%%%%%%%%%%%%%%%%%%%%%%%%%%%%%%%%%%

 \subsection{Scalar production via $\eta$ meson decays}
In this appendix we obtain the branching fraction of an $\eta$ meson to a pion and the scalar particle $\varphi$, 
for the Higgs portal and hadrophilic scalar models discussed in the body of this work. 
Couplings of the scalar $\varphi$ to the $\eta$ meson arise at tree level from its couplings to first-generation quarks Eqns.~\eqref{eq:higgsportalcouplings} or~\eqref{eq:hadrophilicYukawas}, 
and at loop level from its coupling to gluons. 
In the effective theory below a fermion mass threshold and up to naive dimension five,
the couplings to gluons are given by \cite{Gunion:1989we}
\begin{equation}
\label{eq:gluonoperator}
\mathcal{L} \supset  \sum_q \frac{\alpha_S \lambda_{\varphi}^q}{12\pi v} \varphi \, G_{\mu \nu} G^{\mu\nu} \quad ,
\end{equation}
where $\lambda_{\varphi}^q$ are the scalar-quark couplings of~\eqref{eq:higgsportalcouplings} for the minimal Higgs portal model, 
or~\eqref{eq:hadrophilicYukawas} for the hadrophilic scalar model, 
and we sum over quarks $q$ with masses above the effective theory cutoff.
To obtain the couplings to the $\eta$ meson we match the effective theory containing the Yukawa coupling operators Eqns.~\eqref{eq:renL},~\eqref{eq:dim5_yukawas} and the gluon operator Eq.~\eqref{eq:gluonoperator}, 
to the low energy chiral lagrangian. 
Details on the procedure can be found in \cite{Gunion:1989we,Batell:2018fqo}.
From the chiral Lagrangian, 
we find an interaction between our scalar, a pion and an $\eta$ meson given by
\begin{equation}
\nonumber
\mathcal{L} \supset g_{\varphi \eta \pi} \varphi \eta \pi \quad ,
\end{equation}
where
\begin{eqnarray}
\nonumber
 g_{\varphi \eta \pi} &=&
 \bigg[\lambda_\varphi^u-\lambda_\varphi^d 
 + \frac{2}{9}(m_u-m_d)
 \sum_{q=t,b,c}
  \frac{\lambda_\varphi^q}{m_q}
  \,
 \bigg] c_{\varphi \pi \eta} B_0 \quad ,
 \\
 \label{eq:etacoupling}
\end{eqnarray}
and $B_0=m_{\pi^2}/(m_u+m_d)\simeq 2.6 \, \textrm{GeV}$, $c_{\varphi \pi \eta}\simeq 0.82$.
In Eq.~\eqref{eq:etacoupling}, 
the first two terms on the right hand side come from the tree-level Yukawa couplings of the scalar to first-generation quarks,
while the remaining terms come from matching the one-loop scalar-gluon operator.
Using the coupling Eq.~\eqref{eq:etacoupling} we obtain the width of the $\eta$ meson to a scalar and pion,
\begin{eqnarray}
\nonumber
\Gamma_{\eta \rightarrow  \pi^0 \varphi}
&=&
\,
\frac{g_{\varphi \pi^0 \eta}^2}{16 \pi m_\eta^3}
\, \lambda^{1/2}(m_\varphi^2,m_\eta^2,m_{\pi^0}^2) \quad ,
\label{eq:etaSpi}
\end{eqnarray}
where $\lambda$ is the triangle function.
The branching fraction of the $\eta$ meson to a singlet and a pion is calculated using Eq.~\eqref{eq:etaSpi} and the SM width, 
$\Gamma_{\eta}^{SM}=1.31\, \textrm{keV}$ \cite{Patrignani:2016xqp}.
For a scalar with $m_{\varphi} = 200\,\textrm{MeV}$ we find $\br(\eta \to \varphi \pi) = 2\times 10^{-5}\, \sin^2\theta$, which is an order of magnitude larger than the estimate from ref.~\cite{Leutwyler:1989xj}, but is in rough agreement with the estimates from ref.~\cite{Kozlov:1995yd} in the ``I = 1" case.

\bibliographystyle{utphys}
\bibliography{koto}

\end{document}